\documentstyle[12pt]{article}
\input epsf
\begin{document}

\newcommand{\be}{\begin{equation}}
\newcommand{\ee}{\end{equation}}
\def\bq{\begin{eqnarray}}
\def\eq{\end{eqnarray}}

\begin{center}
\large{{\bf QSSC re-examined for the newly discovered SNe Ia}}
\end{center}

\begin{center}
R. G. Vishwakarma,\footnote{E-mail: rvishwa@mate.reduaz.mx}\\
\emph{Department of Mathematics\\
    Autonomous University of Zacatecas\\
  Zacatecas, ZAC C.P. 98060\\
                Mexico}\\

\vspace{0.5cm} 
and\\

\vspace{0.5cm} 
J. V. Narlikar\footnote{E-mail: jvn@iucaa.ernet.in}\\
\emph{Inter-University Centre for Astronomy \& Astrophysics\\
 Post Bag 4, Ganeshkhind, Pune 411 007\\ 
India }
\end{center}

\bigskip
\begin{abstract}\noindent
We examine the possible consistency of the quasi-steady state model with
the newly discovered SNe Ia. The model assumes the existence of metallic
dust ejected from the SNe explosions, which extinguishes light travelling 
over long distances. We find that the model shows a reasonable fit to the 
data, which improves if one takes account of the weak gravitational lensing 
effect of the SNe which have been observed on the brighter side.

\medskip
\noindent
{\it Subject heading:} cosmology: alternative theories, SNe Ia.

\noindent
{\bf Key words:} quasi-steady state cosmology, SNe Ia observations.
\end{abstract}

\newpage
\noindent
{\bf 1. INTRODUCTION}

\noindent
The high redshift supernovae (SNe) Ia explosions look fainter than they are 
expected in the
Einstein-deSitter model, which used to be the favoured
model before these observations were made a few years ago. 
This observed faintness is generally explained by invoking some hypothetical 
source with negative pressure often known as \emph{`dark energy'},
the simplest and the most favoured candidate being a positive 
cosmological constant $\Lambda$.
This happens because the metric distance of an object out to any redshift 
can be increased by incorporating a \emph{`fluid'} with negative
pressure in Einstein's equations.
However, a constant $\Lambda$, is plagued with the so called cosmological
constant problem: why don't we see
the large vacuum energy density
$\rho_{\rm v}\equiv \Lambda/8\pi G \approx 10^{61}$ GeV$^4$,
required to drive inflation in the primordial epochs of the universe
which  is $\approx$$10^{108}$ times larger than
the value required by the SNe observations? 
It is more natural to believe that $\Lambda$ dropped to zero, 
after the inflation was over, rather than try to explain its relic at such a 
small but highly fine-tuned value. Variable $\Lambda$ or `quintessence' models
also fail to solve this problem without proper fine tuning, apart from the
fact that they have a somewhat ad-hoc nature and certainly do not share the
elegance of the overall structure of general relativity.

 An alternative way to explain the faintness of the high redshift SNe Ia
is to consider the absorption of light by metallic dust ejected from the 
SNe explosions$-$an issue which, in the standard approach, is generally 
ignored  while discussing $m$-$z$ relation for SNe Ia. 
However, there is at least one theory, the quasi-steady state 
cosmology (QSSC), which considers this effect as a main ingredient of the
theory. The QSSC is a Machian theory proposed by Hoyle, Burbidge and 
Narlikar (1993, 2000) as an alternative to the Standard Big Bang
Cosmology (SBBC). This cosmology does not have any cosmic epoch when the 
universe was hot and {\it is free from the initial singularity} 
of the SBBC.
It has been shown earlier (Banerjee, et al 2000; Vishwakarma 2002; Narlikar, 
et al 2002) that by taking into account the absorption of SNe light by the 
intergalactic metallic dust, the QSSC explains successfully the SNe Ia data 
from Perlmutter et al (1999), together with SN 1997ff, the highest redshift
SN observed so far (Riess, et al 2001). Extending that work further, we examine
in this paper how well (or badly !) the QSSC fits the new data discovered  with
the Hubble Space Telescope (HST) which include 7 highest redshift SNe Ia known,
all at $z>1.25$ (Riess, et al 2004).

Being based on a Machian theory of gravity, the magnitudes and signs of the 
creation field energy and $\Lambda$ are 
determined by the large scale structure of the universe (Hoyle, et al 1995).
The role of $\Lambda$ (which is negative in this theory) in the dynamics of 
the model is to energize the creation-field by controlling the
expansion of the universe. The repulsive effects (akin to that from a positive
$\Lambda$ in the SBBC) are generated by the creation field which has a 
negative energy density. Therefore the model has cycles of expansion and
contraction (regulated by the creation- and the negative $\Lambda$-fields
respectively) of comparatively shorter period  (around 50 Gyr) superposed 
on a long term (around 1000 Gyr)
steady state-like expansion. Creation of matter is also periodic, being
confined to pockets of strong gravitational fields around compact massive 
objects. These creation centers are `turned on' close to the minimum scale 
size in a typical cycle and are gradually `turned off' during expansion to 
maximum scale size.

The theory seems to meet all the available observational constraints at the
present time. According to the QSSC, the cosmic microwave background (CMB) 
is the relic starlight left by the stars
of the previous cycles which has been thermalized by the metallic whisker 
dust emitted
by the supernovae. It is very interesting to note that {\it the energy 
available from this process is just right to give a radiation background of 
2.7 K at the present epoch} (Hoyle, et al 1994). SBBC, on the other 
hand, does not predict the present temperature of the CMB. 
The theory also explains the observed anisotropy of CMB, including the peaks
at $l\sim 200$ and $l\sim 600$ which are related, in this cosmology, to the
clusters and groups of clusters (Narlikar, et al 2003).
The theory does not face the cosmological constant problem mentioned earlier.
In fact, the $\Lambda$ in the QSSC does not represent
the energy density of the quantum fields, as this model does not experience
the energy scales of quantum gravity except within the local centres of 
creation.

It may also be noted that, unlike SBBC, the QSSC allows the dark matter to be 
baryonic. It may be recalled that the SBBC predicts the existence of 
non-baryonic, though as yet undetected, particles to solve the problems of 
structure formation and of the missing mass in bound gravitational systems 
such as galaxies and clusters of galaxies.  Although there has been a steady 
evolution of views on whether dark matter 
is predominantly cold or hot, there is no satisfactory observational 
evidence for the postulated particles from laboratory physics. 
The predicted density distribution of dark halos which result from N-body 
simulations (Navarro, et al 1996), appears to be inconsistent with 
observations of spiral galaxies (de Blok, et al 2001) or with strong lensing 
in clusters of galaxies (Treu, et al 2003). It is therefore fair to say that 
this scheme has still to demonstrate its viability.
However, in the framework of the QSSC, the dark matter need not be necessarily
non-baryonic. It can be in the form of baryonic matter being the relic of
very old stars of the previous cycles.

For the sake of completeness and for the ready reference, we describe
briefly the mathematical formulation of QSSC in the Appendix. The details of
this cosmology can be found in the papers of Hoyle, Burbidge and Narlikar
mentioned above and in the paper by Sachs et al (1996).

\bigskip
\noindent
{\bf 2. EXTINCTION BY METALLIC DUST}

\noindent
Chitre and Narlikar (1976) were the first to discuss the role of
intergalactic dust in the $m$-$z$ relation, which was however largely ignored 
at that time.
It is, however, generally accepted now that the metallic vapours are ejected 
from the SNe
explosions which are subsequently pushed out of the galaxy through pressure
of shock waves (Hoyle \& Wickramasinghe 1988; Narlikar, et al 1997).
Experiments have shown that metallic vapours on cooling, condense into
elongated whiskers of $\approx$ $0.5-1$ mm length and $\approx$$10^{-6}$ cm 
cross-sectional radius (Donn \& Sears 1963; Nabarro \& Jackson 1958).
Indeed this type of dust extinguishes radiation travelling over long distances
(Aguire 1999; Banerjee, et al 2000; Narlikar, et al 2002; Vishwakarma 2002; 
2003). The density of the dust can be estimated
along the lines of Hoyle, et al (2000). If the metallic whisker production is
taken as 0.1 $M_\odot$ per SN and if the SN production rate is taken as
1 per 30 years per galaxy, the total production per galaxy (of spatial density
$\approx$ 1 per $10^{75}$ cm$^3$) in $10^{10}$ years is $\approx 2/3\times 
10^{41}$ g. The expected whisker density, hence, becomes 
$2/3\times 10^{41}\times 10^{-75}\approx 10^{-34}$ g cm$^{-3}$.
We shall see later that this value is in striking agreement with the
best-fitting value coming from the SNe Ia data.

Earlier work (Banerjee, et al 2000; Vishwakarma 2002; Narlikar, et al 2002) 
has shown that the extinction due to dust adds an extra magnitude 
$\Delta m(z)$ to the apparent magnitude $m(z)$ of a supernova of redshift 
$z$, where
\begin{equation}
\Delta m(z)=1.0857\times ~\kappa ~\rho_{\rm g0}\int_0^z
(1+z')^2\frac{{\rm d}z'}{H(z')}.\label{eq:deltam}
\end{equation}
Here $\kappa$ is the mass absorption coefficient which is effectively constant
over a wide range of wavelengths and is of the order $10^5$ cm$^2$ g$^{-1}$ 
(Wickramasinghe \& Wallis 1996); and 
$\rho_{\rm go}$ is the whisker grain density at the present epoch:
$\rho_{\rm g}~S^3=\rho_{\rm g0}~S_0^3$. 
The net apparent magnitude is then given by
\begin{equation}
 m^{\rm net}(z)=m(z) + \Delta m(z).\label{eq:mnet}
\end{equation}
The usual apparent magnitude $m(z)$ arising from the cosmological evolution 
is given by
\be
m (z) =5 \log[H_0 ~d_{\rm L}(z)] + {\cal M},\label{eq:mageq}
\ee
with the luminosity distance $d_{\rm L}$ given by
\be
d_{\rm L}(z) = (1 + z) \int_0^z \, \frac{{\rm d} z'}{H(z')},
\label{eq:distL}
\ee
for the $k=0$ case of the RW metric. The constant ${\cal M}$ appearing in 
equation (\ref{eq:mageq}) is given by 
${\cal M} \equiv M - 5 \,\log H_0 + constant$, where $M$ is the absolute 
magnitude of the SNe. The Hubble parameter $H(z)$ appearing in equations 
(\ref{eq:deltam}) and (\ref{eq:distL}) is given by (\ref{eq:hubble}).

\begin{figure}[tbh!]
\centerline{{\epsfxsize=14cm {\epsfbox[50 250 550 550]{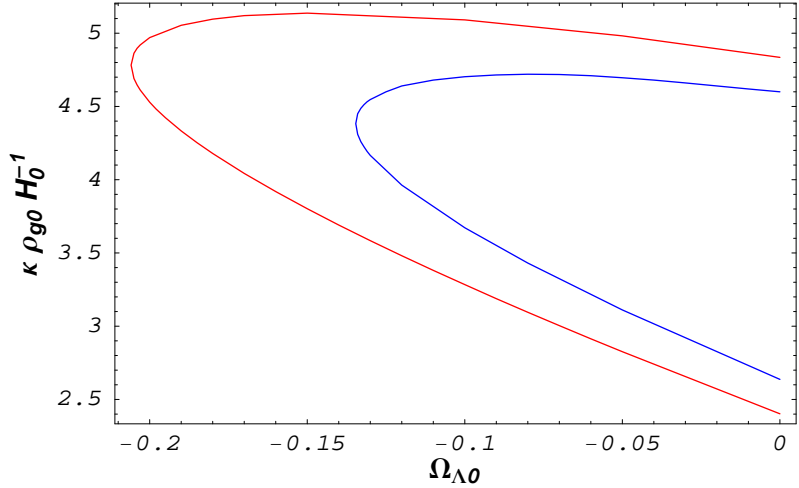}}}}
{\caption{\small The allowed regions by the `gold sample' of SNe Ia data 
(with 157 points) are shown at the 95\% (inner contour) and 99\% 
(outer contour) confidence levels, by marginalizing over ${\cal M}$.
(The parameters $\kappa$, $\rho_{\rm g0}$ and $H_0$ are 
measured in units of $10^5$ cm$^2$ g$^{-1}$, $10^{-34}$ g cm$^{-3}$ and
100 km s$^{-1}$ Mpc$^{-1}$ respectively.)
For an average values $\kappa=5\times 10^5$ cm$^2$ g$^{-1}$  
and $H_0=$ 70 km s$^{-1}$ Mpc$^{-1}$, the estimated range of the whisker
grain density from the data is $3.5\times 10^{-34}$ g cm$^{-3} 
\leq \rho_{\rm g0} \leq 8\times 10^{-34}$ g cm$^{-3}$ at 99\% confidence
level. }}
 \end{figure}

\bigskip
\noindent
{\bf 3. DATA FITTING}

\noindent
We consider the data recently published by Riess et al (2004) which, in 
addition to
having previously observed SNe, also include 16 newly discovered SNe Ia 
by the HST, 6 of them being among the 7 highest redshift SNe Ia known, all at 
redshift $>1.25$. We particularly focus on their `gold sample' of 157 SNe Ia
which is claimed to have a `high confidence' quality of the spectroscopic and
photometric record for individual supernovae. We note that the data points of
this sample are given in terms of distance modulus 
$\mu_o=m^{\rm net}-M=5\log d_{\rm L}+constant$.
However, the zero-point absolute magnitude or
Hubble constant were set arbitrarily for this sample. Therefore, while fitting
the data, we can compare the observed $\mu_o$ with our predicted $m^{\rm net}$
given by equation (\ref{eq:mnet}) and compute $\chi^2$ from
\be
\chi^2 = \sum_{i = 1}^{157} \,\left[ \frac{m^{\rm net}(z_i; ~\Omega_{\Lambda0},
~\kappa \rho_{\rm g0}H_0^{-1}, ~{\cal M}) - \mu_{o, i}}
{\sigma_{\mu_{o, i}}}\right]^2,\label{eq:chi}
\ee
the constant ${\cal M}$ thus playing the role of the normalization constant.
The quantity $\sigma_{\mu_{o, i}}$ is the uncertainty in the distance
modulus $\mu_{o, i}$ of the $i$-th SN.
We consider the simplest QSSC model with $k=0$ case of the
RW metric (\ref{eq:rw}). Thus there are only three independent free parameters
to be estimated from the data: ${\cal M}$, $\kappa \rho_{\rm g0}H_0^{-1}$ and
$\Omega_{\Lambda0}$ (only the last one comes from the field equations, see 
Appendix).
The parameter $\kappa \rho_{\rm g0}H_0^{-1}$, which is 
dimensionless, is of the order of unity if one considers $\kappa$ of the order
$10^5$ cm$^2$ g$^{-1}$, $\rho_{\rm g0}$ of the order $10^{-34}$ g cm$^{-3}$ and
$H_0\sim$ 70 km s$^{-1}$ Mpc$^{-1}$. However, we have kept it as a free 
parameter to be estimated from the data.

By varying the free parameters of the model, we find that $\chi^2$ decreases
as $\Omega_{\Lambda0}$ increases. Thus for the theoretically allowed region
$\Omega_{\Lambda0}<0$, the best-fitting $\chi^2$ is $>201.41$ [at 155 degrees
of freedom (dof), i.e., $\chi^2$/dof = 1.3]. For example, for the models

\medskip
\noindent
$\Omega_{\Lambda0}=-0.1$: $\chi^2$/dof $= 205.85/155=1.33$;\\
$\Omega_{\Lambda0}=-0.2$: $\chi^2$/dof $= 210.36/155=1.36$;\\
$\Omega_{\Lambda0}=-0.3$: $\chi^2$/dof $= 214.91/155=1.39$.

\medskip
\noindent
In order to compare these results with the SBBC, we note that for a constant
$\Lambda$, the best-fitting flat model ($\Omega_{\rm m0}+\Omega_{\Lambda0}=1)$
and the global best-fitting model (without any such constraint) are obtained as

\medskip
\noindent
$\Omega_{\Lambda0}=1-\Omega_{\rm m0}=0.69$, with  $\chi^2$/dof $= 177.07/155
=1.14$;\\
$\Omega_{\Lambda0}=0.98$, $\Omega_{\rm m0}=0.46$, with  $\chi^2$/dof 
$= 175.04/154=1.14$.

\medskip
\noindent
The so called `\emph{concordance}' model
$\Omega_{\rm m0}=1-\Omega_{\Lambda0}=0.27$ 
gives  $\chi^2=178.17$, which is slightly higher than the best-fitting
value of $\chi^2$ for the flat model mentioned above. In fact, there is an 
almost flat valley around $\Omega_{\rm m0}=0.3$ on the 
$\Omega_{\rm m0}$-${\cal M}$-$\chi^2$ 
surface where $\chi^2$ does not vary significantly and hover around 177-178. 
For example, the models $\Omega_{\rm m0}(=1-\Omega_{\Lambda0})=0.28$ and 0.3,
respectively, give $\chi^2=177.67$ and 177.13.

Though the fit in the QSSC is certainly not as good as in the SBBC, it
is by no means rejectable. Moreover the estimated value of
$\kappa \rho_{\rm g0}H_0^{-1}$ is indeed of the order of unity:
the best-fitting values of $\kappa \rho_{\rm g0}H_0^{-1}$ for the 
cases $\Omega_{\Lambda0}=$ -0.1, -0.2 and -0.3 are respectively 4.19, 4.75 
and 5.30, as expected from the theory.
Here, the parameters $\kappa$, $\rho_{\rm g0}$ and $H_0$ are 
measured in units of $10^5$ cm$^2$ g$^{-1}$, $10^{-34}$ g cm$^{-3}$ and
100 km s$^{-1}$ Mpc$^{-1}$ respectively. 
In Figure 1, we have shown the allowed regions in the parameter space 
 $\Omega_{\Lambda0}-\kappa \rho_{\rm g0}H_0^{-1}$ at 95\% and 99\% confidence
levels, by marginalizing over the parameter ${\cal M}$. Figure 2 
compares the best-fitting theoretical models with the actual
data points.

\begin{figure}[tbh!]
\centerline{{\epsfxsize=14cm {\epsfbox[50 250 550 550]{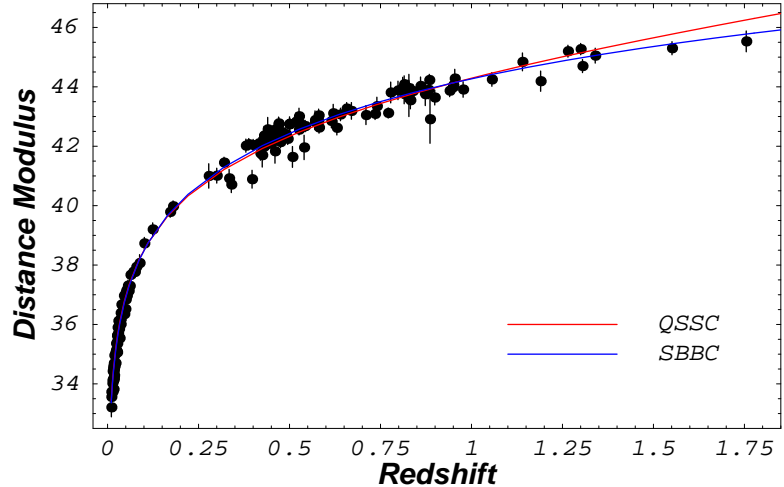}}}}
{\caption{\small Best-fitting flat models in QSSC and SBBC are compared with
the `gold sample' of SNe Ia data from Riess et al (2004). The QSSC model 
corresponds to $\Omega_{\Lambda0}=-0.2$.}}
 \end{figure}

\bigskip
\noindent
{\bf 4. EFFECTS OF WEAK LENSING}

\noindent
Weak gravitational lensing is an unavoidable systematic uncertainty in the use
of SNe Ia as standard candles. As the universe is inhomogeneous in matter
distribution, the SNe fluxes are magnified by foreground galaxy excess and
demagnified by foreground galaxy deficit, compared to a smooth matter 
distribution. Recently Williams \& Song (2004) have reported such a correlation
between the magnitudes of 55 SNe from the sample of Tonry et al (2003) and
foreground galaxy overdensity. They have found the difference between the
most magnified and the most demagnified SNe as about 0.3-0.4 mag. Wang (2004)
has claimed further evidence of gravitational magnification of three brightest
SNe from the Riess et al (2004) sample:

\medskip
\noindent
SN1997as ($z=0.508$, $\mu_o=41.64$): magnified by $2.10\pm0.68$;\\
SN2000eg ($z=0.540$, $\mu_o=41.96$): magnified by $1.80\pm0.70$;\\
SN1997as ($z=0.886$, $\mu_o=42.91$): magnified by $2.42\pm1.98$.

\medskip
\noindent
These high magnification factors $\sim 2$ from weak lensing are though somewhat
surprising, as also mentioned by Menard \& Dalal (2004) who claim not to find
any significant correlation between SN magnification and foreground galaxy 
overdensity. However, even if we assume a mild magnification of the 
above-mentioned SNe just by an average 0.5 mag, this improves the fit
considerably:

\medskip
\noindent
$\Omega_{\Lambda0}=-0.1$: $\chi^2_{\rm improved}$/dof $= 198.55/155=1.28$;\\
$\Omega_{\Lambda0}=-0.2$: $\chi^2_{\rm improved}$/dof $= 203.17/155=1.31$;\\
$\Omega_{\Lambda0}=-0.3$: $\chi^2_{\rm improved}$/dof $= 207.82/155=1.34$.

\medskip
\noindent
Though the weak lensing effects are estimated to be small for SNe at $z<1$,
they are non-negligible for higher redshift SNe. 
As more SNe are discovered at higher redshifts, it becomes increasingly
important to minimize the effect of weak lensing, say, by considering 
the flux-averaging (Wang, 2000).

\bigskip
\noindent
{\bf 5. CONCLUDING REMARKS}

\noindent
The redshift magnitude test has had a chequered history. During the 1960s and 
the 1970s it was used to draw very categorical conclusions.  The deceleration 
parameter $q_0$ was then claimed to lie between 0 and 1 and thus it was claimed
that the universe is decelerating.  Gunn and Oke (1975), however, pointed out
that there remained observational errors to be allowed for, that vitiated that
conclusion. For example, aperture correction, luminosity evolution, etc. were 
to be allowed for.  It was then realized that the test is not as conclusive 
in selecting a cosmological model as it earlier appeared to be.

Today's situation, we feel, is hardly different. There has been considerable 
progress in our understanding of the physics of supernovae, yet it is hard 
to imagine that the peak luminosity of Type Ia supernovae remains a standard 
candle over a redshift exceeding 1. Evolution has long been assumed in various
other cosmological tests, like the counts of galaxies and radio sources, the 
variation of angular size with redshift, etc. The assumption of a non-evolving
standard candle for supernovae therefore needs to be more critically examined 
than has been hitherto.

    The possible role of gravitational lensing in amplifying the 
supernova luminosity at high redshifts has been discussed by several authors 
and we have applied those ideas here to illustrate the difference it can make 
to any conclusion drawn from the data.  Additionally, the role of 
intergalactic dust still remains to be appreciated fully and we have 
demonstrated here the possible difference it can make to the viability of 
a model.

  Contrary to the widespread belief that these caveats do not matter or have 
already been allowed for, we retain a healthy skepticism of this test as 
contributing to a `precise' determination of cosmological parameters.  For 
this reason we are satisfied with the level of `goodness of fit' obtained 
here for the QSSC.  The fit could no doubt be improved by tinkering with the 
parameters; but given the observational uncertainties, we do not feel it 
worthwhile to undertake that exercise.

\bigskip
\noindent
{\bf ACKNOWLEDGEMENTS}

\noindent
JVN thanks the College de France for hospitality and RGV thanks the Abdus 
Salam ICTP for hospitality received under the associateship programme 
during the course of this work. The authors also thank Riess et al for 
providing their gold sample of data.

\newpage
\begin{center}
{\bf APPENDIX}
\end{center}

\setcounter{equation}{0}
\renewcommand{\theequation}{A.\arabic{equation}}

\noindent
The field equations of QSSC, which arise from a Machian theory of gravity,
are more general than the Einstein field equations:
\begin{equation}
R^{ij}-\frac{1}{2} R ~ g^{ij} - \Lambda  ~ g^{ij}
=-8\pi G\left[T^{ij}_{\rm matter} +T^{ij}_{\rm creation} 
\right].\label{eq:feq}
\end{equation}
Here the speed of light is taken as unity. The first term on the right is 
the usual energy momentum tensor of matter
\begin{equation}
T^{ij}_{\rm matter}=(\rho+p)u^i ~u^j+p ~g^{ij},\label{eq:tem}
\end{equation}
whereas the second term denotes the contribution from a trace-free zero rest 
mass scalar field $c$ of \emph{negative} energy and stresses with gradient
$c_i\equiv \partial c/\partial x^i$:
\be
T^{ij}_{\rm creation}= -f\left(c^ic^j+\frac{1}{4}c^\ell c_\ell ~ g^{ij}\right)\label{eq:tc},
\ee
with a positive coupling constant $f$. In the case of the homogeneous 
isotropic spacetime described by the RW metric 
\be
{\rm d}s^2=-{\rm d}t^2 + S^2(t) \left\{\frac{{\rm d}r^2}{1-kr^2} +
r^2({\rm d}\theta^2+\sin^2\theta ~{\rm d}\phi^2)\right\},\label{eq:rw}
\ee
the field equations (\ref{eq:feq}$-$\ref{eq:tc})
lead to the following two equations for the `dust' universe ($p=0$):
\begin{equation}
\frac{\dot S^2}{S^2}+\frac{k}{S^2}=\frac{\Lambda}{3}+\frac{8\pi
G}{3}(\rho _{\rm m}+\rho _{\rm c}), ~ ~ ~
\rho _{\rm c}\equiv-\frac{3}{4}f\dot c^2;\label{eq:fman} 
\end{equation}
\begin{equation}
2\frac{\ddot S}{S}+\frac{\dot S^2}{S^2}+\frac{k}{S^2}=\Lambda-8\pi G p_{\rm c},
~ ~ ~ p_{\rm c}\equiv-\frac{1}{4}f\dot c^2.\label{eq:ray}
\end{equation}
By assuming that the present epoch is 
represented by the non-creative mode of the model, i. e.,
$T^{ij}_{{\rm matter} ~;j}=T^{ij}_{{\rm creation} ~;j}=0$, giving 
\be
\frac{8\pi G}{3}\rho_{\rm m}=\frac{A}{S^3}, ~ ~
\frac{8\pi G}{3}\rho_{\rm c}=\frac{B}{S^4}, ~ ~ ~ A, B=\mbox{constants},\label{eq:ab}
\ee
equation
(\ref{eq:fman}) can be solved to give the scale factor in the form
\begin{equation}
S=\bar S[1+\eta \cos \psi(t)].\label{eq:sfactor} 
\end{equation}
The parameter $\eta$ lies in the range $0<\eta<1$ 
and the function $\psi$ is given by
\begin{equation}
\dot\psi^2=-\frac{\Lambda}{3}(1+\eta\cos\psi)^{-2}\{6+4\eta\cos\psi+\eta^2(1+
\cos^2\psi)\}.\label{eq:psy}
\end{equation}
Obviously $\Lambda$ is negative, as has been mentioned earlier. It is also
obvious from equation  (\ref{eq:sfactor}) that $S$ never becomes zero and 
oscillates between
\begin{equation}
\bar S(1-\eta)\equiv S_{\rm min}\leq S\leq S_{\rm max} \equiv \bar S(1+\eta).\label{eq:zmax}
\end{equation}
The constant 
$\bar S$, appearing in (\ref{eq:sfactor}) and (\ref{eq:zmax}), is given by
\be
A=2k\bar S-\frac{4}{3}\Lambda {\bar S}^3 (1+\eta^2),\label{eq:Cbar1}
\ee
\be
B=k {\bar S}^2 (1-\eta^2)-\frac{1}{3}\Lambda {\bar S}^4 (1-\eta^2)(3+\eta^2),\label{eq:Cbar2}
\ee
which can be obtained from equations (\ref{eq:fman}) and (\ref{eq:ray}) using
(\ref{eq:ab}) and (\ref{eq:sfactor}) therein. Equations (\ref{eq:Cbar1}) and
(\ref{eq:Cbar2}) can be recast in the following forms in terms of the 
different energy components computed at the present epoch:
\be
\Omega_{\rm m0}=2\Omega_{k0}(1+\bar z)^{-1}-4\Omega_{\Lambda0}(1+\bar z)^{-3}(1+\eta^2),\label{eq:a1}
\ee
\be
\Omega_{\rm c0}=-\Omega_{k0}(1+\bar z)^{-2} (1-\eta^2)-4\Omega_{\Lambda0}(1+\bar z)^{-4}(1-\eta^2)(3+\eta^2),\label{eq:a2}
\ee
where, as usual,
\be
\Omega_{\rm m0}\equiv \frac{8\pi G}{3H^2_0}\rho_{\rm m0}, ~ ~ 
\Omega_{k0}\equiv \frac{k}{H^2_0 S^2_0},  ~ ~ 
\Omega_{\Lambda0}\equiv \frac{\Lambda}{3H^2_0},  ~ ~ 
\Omega_{\rm c}\equiv \frac{8\pi G}{3H^2_0}\rho_{\rm c0}.
\ee
In terms of these dimensionless parameters, equations (\ref{eq:fman}) and 
(\ref{eq:ray}), by the use of (\ref{eq:ab}), reduce to
\begin{equation}
H(z)=H_0[\Omega_{\Lambda0}-\Omega_{k0}(1+z)^2+\Omega_{\rm m0}(1+z)^3+
\Omega_{c0}(1+z)^4]^{1/2},\label{eq:hubble}
\end{equation}
\begin{equation}
2 q(z)= \left[\frac{H_0}{H(z)}\right]^2[\Omega_{\rm m0}(1+z)^3-2\Omega_{\Lambda0}
+2\Omega_{c0}(1+z)^4].\label{eq:decel}
\end{equation}
It should be noted that not all the parameters, introduced above, are 
independent. For example, equation (\ref{eq:hubble}) suggests that
\be
\Omega_{\Lambda0}-\Omega_{k0}+\Omega_{\rm m0}+\Omega_{\rm c0}=1.
\ee
This equation also suggests that at the maximum redshift $z_{\rm max}$ 
(say, in the present cycle), one has the following identity:
\begin{equation}
\Omega_{\Lambda0}-\Omega_{k0}(1+z_{\rm max})^2
+\Omega_{\rm m0}(1+z_{\rm max})^3+\Omega_{c0}(1+z_{\rm max})^4=0.
\end{equation}
Also equation (\ref{eq:zmax}) suggests that  $z_{\rm max}$,
$\bar z$ and $\eta$ are related by
\be
1+\bar z \equiv \frac{S_0}{\bar S}=(1-\eta)(1+z_{\rm max}).
\ee
Thus out the 7 parameters $\Omega_{\Lambda0}$, $\Omega_{\rm m0}$, 
$\Omega_{\rm c0}$,
$\Omega_{k0}$, $\bar z$, $\eta$ and $z_{\rm max}$, only 3 parameters, say, 
$\Omega_{k0}$, $\Omega_{\Lambda0}$ and $z_{\rm max}$ are independent. For
the case $k=0$, which gives the simplest one of the QSSC models, only two 
independent parameters $\Omega_{\Lambda0}$ and $z_{\rm max}$ are left out. 
We further consider $z_{\rm max}=5$, as has been done in the earlier papers 
on the QSSC, which leaves only one free parameter, say, $\Omega_{\Lambda0}$ 
coming from the field equations.

\bigskip
\noindent
{\bf REFERENCES}\\
Aguire A. N., 1999, ApJ, 512, L19\\
Banerjee S. K., Narlikar J. V., Wickramasinghe N. C., Hoyle F., Burbidge,

\hspace{.5cm} G., 2000, ApJ, 119, 2583

\noindent
de Blok, W. J. G et al, 2001, Astron. J., 122, 2396\\
Chitre, S. M., Narlikar J. V., 1976, Astrophys. Space Sc., 44, 101\\
Donn B., Sears G. W., 1963, Science, 140, 1208\\
Gunn J. E., Oke J. B., 1975, ApJ, 195, 255\\
Hoyle F., Wickramasinghe N. C., 1988, Astrophys. Space Sc. 147, 245\\
Hoyle F., Burbidge G., Narlikar J. V., 1993, ApJ, 410, 437\\
Hoyle F., Burbidge G., Narlikar J. V., 1994, MNRAS, 267, 1007\\
Hoyle F., Burbidge G., Narlikar J. V., 1995, Proc. R. Soc. London A, 448,

\hspace{.5cm}  191

\noindent
Hoyle F., Burbidge G., Narlikar J. V., 2000, {\it A Different Approach to

\hspace{.5cm}       Cosmology}, (Cambridge: Cambridge Univ. Press)

\noindent
Menard B., Dalal N., 2004, preprint, astro-ph/0407023\\
Nabarro F. R. N., Jackson P. J., 1958, in \emph{Growth and Perfection in Crystals}, 

\hspace{.5cm} eds. R. H. Duramus, et al, (J. Wiley, New York)

\noindent
Narlikar J. V., Wickramasinghe N. C., Sachs R., Hoyle F., 1997, Int. J. Mod.

\hspace{.5cm}     Phys. D, 6, 125

\noindent
Narlikar J. V., Vishwakarma, R. G., Burbidge G., 2002, PASP, 114, 1092\\
Narlikar J. V., Vishwakarma, R. G., Hajian A., Souradeep T., Burbidge G., 

\hspace{.5cm} Hoyle F., 2003, ApJ, 585, 1

\noindent
Navarro J. F., et al, 1996, ApJ, 462, 563\\
Perlmutter S., et al., 1999, ApJ., 517, 565\\
Riess A. G., et al., 2001, ApJ., 560, 49\\
Riess A. G., et al., 2004, ApJ., 607, 665\\
Sachs R., Narlikar J. V., Hoyle F., 1996, A\&A, 313, 703\\
Tonry J. L., et al, 2003, preprint, astro-ph/0305008\\
Treu T. et al, 2003, preprint, astro-ph/0311052\\
Vishwakarma R. G.,  2002, MNRAS, 331, 776\\
Vishwakarma, R. G., 2003, MNRAS, 345, 545\\
Wang Y., 2000, ApJ., 536, 531\\
Wang Y., 2004, preprint, astro-ph/0406635\\
Wickramasinghe N. C., Wallis D. H., 1996, Astrophys. Space Sc.

\hspace{.5cm} 240, 157

\noindent
Williams L. L. R., Song J., 2004, preprint, astro-ph/0403680\\

\end{document}